\documentclass[oldversion]{aa}
\usepackage{txfonts}
\usepackage{graphicx}

\usepackage{natbib}
\bibpunct{(}{)}{;}{a}{}{,} 

\newcommand{\corot}{\emph{CoRoT}}
\newcommand{\exoseven}{CoRoT-7}

\begin{document}

\title{Ground-based photometry of space-based transit detections: Photometric follow-up of the \corot \thanks{The \corot\ space mission, launched on December 27th 2006, has been developed and is operated by CNES, with the contribution of Austria, Belgium, Brasil, ESA (RSSD and Science Program), Germany and Spain.} mission }
\author
{H.J. Deeg \inst{1} \and M. Gillon \inst{2,3}  \and A. Shporer \inst{4} \and D. Rouan \inst{5}  \and B. Stecklum\inst{6} \and S. Aigrain \inst{7} \and A. Alapini \inst{7} \and J.M. Almenara \inst{1}    \and R. Alonso   \inst{8} \and M. Barbieri \inst{8} \and F. Bouchy  \inst{9} \and J. Eisl\"offel \inst{6} \and A. Erikson \inst{10} \and M. Fridlund \inst{11} \and P. Eigm\"uller  \inst{6}\and G. Handler \inst{12} \and A. Hatzes \inst{6} \and P. Kabath \inst{10}\and M. Lendl \inst{12} \and T. Mazeh \inst{4} \and C. Moutou \inst{8} \and D. Queloz \inst{2} \and H. Rauer \inst{10,13} \and M. Rabus \inst{1}\and B. Tingley\inst{1} \and R. Titz\inst{10}}
\offprints
{H.J. Deeg}

\institute{
Instituto de Astrof\'\i sica de Canarias, C. Via Lactea S/N, 38205 La Laguna, Tenerife, Spain 
\and Observatoire de Gen\`eve, Universit/«e de Gen\`eve, 51 chemin des Maillettes, 1290 Sauverny, Switzerland 
\and Institut d'Astrophysique et de GŽophysique, UniversitŽ de Lige, 4000 Lige, Belgium
\and School of Physics and Astronomy, Tel Aviv University, Tel Aviv 69978, Israel 
\and LESIA, Observatoire de Paris-Meudon, 5 place Jules Janssen, 92195 Meudon, France 
\and Th\"uringer Landessternwarte, Sternwarte 5, Tautenburg 5, D-07778 Tautenburg, Germany 
\and School of Physics, University of Exeter, Stocker Road, Exeter EX4 4QL, United Kingdom 
\and Laboratoire d'Astrophysique de Marseille, 38, rue Fr\'{e}d\'{e}ric Joliot-Curie, 13388 Marseille cedex 13
\and Institut d' Astrophysique de Paris, Universit/«e Pierre \& Marie Curie, 98bis Bd Arago, 75014 Paris, France 
\and Institute of Planetary Research, DLR, Rutherfordstr. 2, D-12489 Berlin, Germany 
\and Research and Scientific Support Department, ESTEC/ESA, 2200 Noordwijk, The Netherlands 
\and Institute of Astronomy, University of Vienna, T\"urkenschanzstr. 17, A-1180 Vienna, Austria 
\and Center for Astronomy and Astrophysics, TU Berlin, Hardenbergstr. 36, 10623 Berlin 
}

\date
{Received ....; accepted ....}



\abstract {The motivation, techniques and performance of the ground-based photometric follow-up of transit detections by the \corot\ space mission are presented. Its principal raison d'\^{e}tre arises from the much higher spatial resolution of common ground-based telescopes in comparison to \corot's cameras. This allows the identification of many transit candidates as arising from eclipsing binaries that are contaminating \corot's lightcurves, even in low-amplitude transit events that cannot be detected with ground-based obervations. For the ground observations, 'on'-'off' photometry is now largely employed, in which only a short timeseries during a transit and a section outside a transit is observed and compared photometrically. \corot planet candidates' transits are being observed by a dedicated team with access to telescopes with sizes ranging from 0.2 to 2 m. As an example, the process that led to the rejection of contaminating eclipsing binaries near the host star of the Super-Earth planet \exoseven b is shown. Experiences and techniques from this work may also be useful for other transit-detection experiments, when the discovery instrument obtains data with a relatively low angular resolution.}

\authorrunning{Deeg et al.}
\titlerunning{Ground-based photometry of space-based transit detections}

\maketitle

\section{Introduction}
The launch of the \corot\ ({\it Co}nvection, {\it Ro}tation \& {\it T}ransits) spacecraft at the end of 2006 opened a new era in the search for transiting planets. This first instrument designed to detect exoplanets from space  \citep{baglin+07, corotbook06, garrido+06} has now obtained about two years of scientific observations with all instrumental parameters fully within the objectives, and its first planet detections have been reported by \citet{barge+08}, \citet{alonso+08a}, \citet{aigrain+08} and \citet{deleuil+08}. Technically, \corot\ is a space telescope of 27 cm $\diameter$ with a CCD camera that has a field of view of 1.5 $\deg$ by 3 $\deg$ for the detection of transiting planets; another field of similar size is reserved for asteroseismologic observations. In observing runs lasting up to 150 days, \corot's exoplanet search acquires simultaneous lightcurves for about 12\ 000 stars, which are then analysed for the presence of transit-like events.   

While the \corot\ spacecraft will deliver the high-precision photometry that is central to this project, the scientific output of this mission is strongly dependent on a dedicated ground-based  observing program. This program began prior to the launch, with observations for the selection of the sample fields and for the characterisation of the targets therein; described by 
\citet{Deleuil+06, Deleuil+09}. The current inflight follow-up is centered on the rejection of false alarms among planet-candidates found by the spacecraft, on the verification of candidates as planets, and on the characterisation of verified planets in more detail. The secure rejection of false alarms and --if possible-- the positive verification of a candidate's planetary nature is essential to achieve the mission's primary scientific goal of delivering reliable exoplanet detections. For the candidate verification process, a multi-step follow-up strategy is used that follows the scheme outlined by \citet{adb+04} which is designed to save time, cost, and effort. It begins with relatively simple tests (e.g. detailed analysis of \corot's lightcurves; analysis of snapshots of candidate fields; transit-observations on small instruments) and for surviving candidates progresses to more complicated and expensive ones (in terms of telescope time, instrumental requirements, and efforts in data-analysis). Photometric follow-up observations form an integral part of this approach and are complimentary to low- and high-resolution spectroscopy undertaken for stellar classification and radial velocity 
measurements. In the mission's preparatory phase, the organizational structure was laid for two working groups, one on spectroscopy and one on photometry. The later one's work is described in the following sections, giving the motivation for the photometric follow-up observations, the techniques used and some examples of the results obtained. 

\section{The motivation for ground-based follow-up}

\begin{figure*}
   \centering
    \includegraphics[width=15cm]{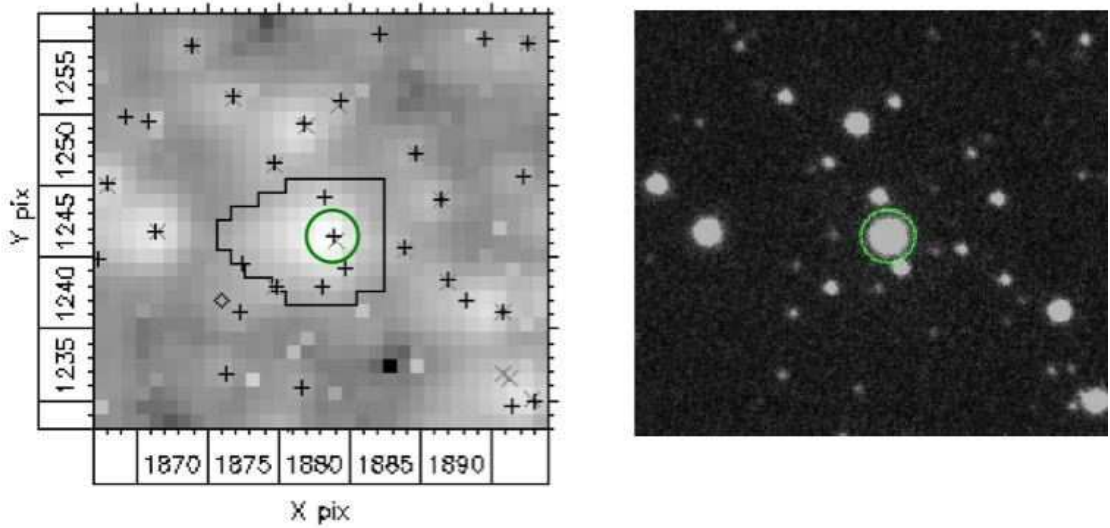}
  
   \caption{Left: Section of a \corot\ target field around a target star (LRc01\_E1\_2376, marked with a circle) as obtained by the satellite. The irregular shape around it indicates the aperture from which photometry is extracted. All stars contained in the Exo-Dat data base are marked by crosses -- several contaminating stars fall within the target aperture. The image size is about 1$\arcmin$. Right: The same field acquired with the CCD camera of the IAC 80cm telescope in moderate (1.5$\arcsec$) seeing. }
              \label{fig:corotwin_vs_iac80}
    \end{figure*}

At the onset, it may appear counterintuitive that ground-based photometric follow-up may be of any use, considering the generally much higher photometric precision of the space-based discovery instrument. The essence of the ground-based follow-up photometry arises from the higher angular resolution achievable compared to photometry from the discovery instrument. In the case of \corot\, its optical resolution is rather poor, with a PSF that contains 50\% of its flux in an elongated shape of about 35$\arcsec$ x 23$\arcsec$ size\footnote{The PSF's irregular shape is due to the dispersion device in the \corot\ lightpath; size and shape depend also on the stellar color and the position in the focal plane} and photometry is measured through apertures of corresponding sizes (Fig.~\ref{fig:corotwin_vs_iac80}). As will be shown in later sections, the increase in spatial resolution from ground-based instruments allows the identification of false alarms even if the originally detected transit-event cannot be reproduced due to the much lower photometric precision from the ground.

There are several sources of false alarms in the detection of transiting planets, mostly related to eclipsing binary (EB) stars; for an overview see  \citet{brown03}. Many of these can be identified from a detailed analysis of the discovery lightcurves in combination with knowledge about the target stars' stellar types, based on indicators like the duration, depth, shape or color of transits, or from the presence of off-eclipse variability or faint secondary transits; e.g.  \citet{eisloeffel04, barge+09}. Of particular worry are faint EBs close to a target star, whose light falls 
within the aperture of the target. These contaminating eclipsing binaries (CEB) may generate transit-like signals that are photometrically undistinguishable from planetary transits. Estimations by \citet{brown03} show that CEBs are the most frequent source of false alarms in ground-based transit searches. Considering the large PSF of  \corot\, the dominance of this source of false alarms may be even more pronounced. This motivates the \corot\ ground-based photometric  follow-up program, which is consequently geared towards the detection of CEBs. Its principles, which are valid for any discovery experiment with initial photometry based on data of relatively low angular resolution, are introduced in more detail in the next section.

\section{Technical considerations for the photometric identification of false alarms}
\subsection{Resolving contaminating sources in images of higher resolution}
In the following, we consider as the \emph{target} the star that has been intended to be observed in a given photometric lightcurve. The probability $P_{\mathit{detc}}$ to detect a contaminating source through observations with a higher angular resolution may be approximated by the expression:
\begin{equation}
P_{\mathit{detc}}=1 - \frac{A_\mathit{psf,hi}}{A_\mathit{psf,lo}} 
\label{eq:pdetc}
\end{equation}
where  $A_\mathit{psf,hi}$ and ${A_\mathit{psf,lo}}$ are the areas of the PSF in the high and low resolution imaging, respectively. This equation is based on the assumption that the contaminating object has no physical relation with the target  and that its position relative to the target is at random. Then, {for example, a reduction of the PSF size from CoRoT's 630 arcsec$^2$ to one of 1.1 arcsec$^2$ (corresponding to a PSF with a diameter of 1.2\arcsec) leads to the expectation that 99.8\% } of the contaminating stars will be found as separate sources in the higher-resolution imaging. Photometric follow-up with higher angular resolution than the discovery instrument is hence able to resolve most cases of signals from unbound contaminating stars that appear merged into the time-series obtained by the low-resolution  discovery instrument. 

We note that the remaining inability of the photometric follow-up to resolve extremely close contaminants, {or to resolve contaminants that are in a physical triple with the target}, is complementary to spectroscopic observations, which can generally detect and distinguish the signals of contaminants close enough to fall within the spectrometer's entry aperture. The detection of contaminants at larger target distances is however better done with photometry, since spectroscopic observations will not show any indication of the false alarm source in such cases. In the particular case of searches for small planets, a spectroscopic non-detection of radial velocity variations presents an ambigous result as it may also indicate that there is a planet, but with a mass too small to be detectable by the instrument; additional photometric follow-up may therefore resolve this ambiguity. 

\subsection{The identification of contaminating variable stars\label{identification}}
Let us assume that the discovery instrument reports a transit-like signal with a variation\footnote{In this work, we consider both the flux-variation from a transit-signal and the corresponding magnitude variation as values that are always positive; hence for small values of $\Delta F/F <<1$, the conversion to magnitudes is given by $\Delta m = -2.5\log(1-\Delta F/F)\approx1.087\Delta F/F$}  of $(\Delta F/F)_s$. The flux $F_s$ reported by the instrument is given by the sum:
\begin{equation}
F_s= g \left( k_t F_t + \sum{k_i F_i} \right)
\label{eq:fluxsum}
\end{equation}
where $g$ is an instrument gain, $F_t$ is the flux of the target star; the $F_i$ are the fluxes of contaminating stars, and the 'confusion factors' $k_t$ and $k_i$ are the fractions of light in the stellar PSF which fall into the given photometric aperture.

A positive detection by the discovery instrument is then indicated if its signal corresponds to a brightness variation $\Delta F_t/F_t$ that is intrinsic to the target, that is if
\begin{equation}
 (\Delta F/F)_s = \frac{k_t \Delta F_t}{k_t F_t + \sum{k_i F_i}} \approx {(\Delta F/F)}_t\ 
\label{eq:positive_case}
\end{equation}
where the right-hand approximation assumes a contaminating flux that is small compared to the target flux, e.g. $\sum{k_i F_i}<< F_t$. The negative case or 'false alarm condition' is then given if the following holds, where one of the contaminating stars (denoted with subscript $c$, where $c \in i$) generates the discovery signal: 
 \begin{equation}
 (\Delta F/F)_s = \frac{k_c \Delta F_c}{k_t F_t + \sum{k_i F_i}} 
\label{eq:neg_case}
\end{equation}

\noindent Follow-up photometry may now do two things: \\
-- It may attempt to {\it verify the positive case} of Eq.~\ref{eq:positive_case}, by reproducing the discovery signal, showing that this signal truly arises from the target.  In transit surveys of very high photometric precision, such as in space missions, low amplitude transit signals are however also be discovered, and a reliable reproduction of these transits in ground-based follow-up will often be impossible.\\
-- Ground-based follow-up may attempt to {\it falsify the negative case} of Eq.~\ref{eq:neg_case}, showing that none of the contaminating stars is generating the discovery signal  $(\Delta F/F)_s$. Typically, the follow-up will show that none of the contaminating stars exhibit brightness variations that are sufficiently strong to be the cause of a false alarm.

 In the following, the task of 'falsifying the false alarm condition' is being developed in more detail. For simplicity, we consider only one contaminating star and also set $k_t \approx 1$, meaning that the target's PSF falls almost entirely into the photometric aperture. 

For a contaminating star with $F_c\, k_c << F_t$, meaning that it contributes only with a small fraction to the light falling into the aperture, we may then rewrite the false alarm condition as:
\begin{equation}
 \frac{\Delta F_c }{F_c} \approx  \left(\frac{F_t}{F_c}\right)\,k_c^{-1}\,(\Delta F/F)_s \ .
 \label{eq:facond}
\end{equation}
If the confusion factor $k_c$ is unkown, a modified condition useful to identifiy {\it potential} false alarms may also be derived by setting  $k_c = 1$; these alarms are then identified if:
\begin{equation}
 \frac{\Delta F_c }{F_c} \ga  \left(\frac{F_t}{F_c}\right)\,(\Delta F/F)_s \ .
 \label{eq:facondk1}
\end{equation}
This modified false alarm condition is useful for a first search for contaminating variable stars. It will identify all of them; it may however also flag contaminating stars whose brightness variation is really too weak to be the source for a false alarm. For contaminants identified by this condition a more detailed analysis is then needed; e.g. by testing against the previous Eq.~\ref{eq:facond}.

In order to avoid separate equations for the cases of known or unknown $k_c$, an {\it unknown} $k_c$ should be  replaced by $k_c=1$ in the remainder of this section, whereas  for a {\it known} $k_c$ the correspondences $\ga or \la$ should be replaced by '$\approx$'.

After conversion to magnitudes, the false alarm conditions of Eqs.~\ref{eq:facond} and \ref{eq:facondk1} are given by
\begin{equation}
\Delta m_c \ga -2.5 \log\left(10^{- 0.4 m_{\mathit{diff}}} -k^{-1}\,(\Delta F/F)_s\right) -m_{\mathit{diff}}
\label{eq:fa_mag}
\end{equation}
where $\Delta m_c$ corresponds to the magnitude variation of the contaminating star and $m_{\mathit{diff}}$ is the difference of (target -- contaminant) magnitudes. For a full expression of Eq.~\ref{eq:fa_mag} in magnitudes, we may replace\ $(\Delta F/F)_s$ by $0.921\, \Delta m_s$, with $m_s$ being the magnitude variation reported by the discovery instrument. 

The above relations show that testing of even very weak transit candidates using ground-based instruments is possible. {As an example (based on LRc02\_E1\_0483 in Table~\ref{tab:ceb}), consider a discovery signal with $(\Delta F/F)_s = 1/1000$ or $\Delta m_c \approx 1$ mmag and a contaminating star with $k_c = 0.5$ that is 5 magnitudes fainter than the target, respectively ${F_t}/{F_c} = 100$. If the eclipse signal originates in the contaminating star and following Eq.~\ref{eq:facond}, its brightness variation would have to be $\Delta F_c/F_c = 0.2$, which is typically easily verifyable from the ground. On the other hand, if the signal is arising from the target, its 1 mmag signal could be detected by ground-based photometry only unreliably; however, the absence of a strong signal in any contaminating star would allow the deduction that the target is the source.}

The need for a positive argument in the logarithmic term in Eq.~\ref{eq:fa_mag} corresponds to the obvious fact that any contaminating star needs to generate a flux in the aperture that at least corresponds to the observed flux variation; e.g. that $F_c \,k_c \ge \Delta F_s/g$, or, in magnitudes, that a contaminating star needs to fulfill 
\begin{equation}
m_{\mathit{diff}} \le -2.5\log\left((\Delta F/F)_s\, k^{-1}\right)\ .
\label{eq:maxconftheo}
\end{equation}
A contaminating star with the minimum brightness would, however, need to turn its flux completely on and off\footnote{Eq.~\ref{eq:maxconftheo} is also identical to setting $\Delta F_c = F_c$ in Eq.~\ref{eq:facond}, which is the maximum possible brightness variation of a contamination source, and converting to magnitudes} in order to generate the observed $(\Delta F/F)_s$.  If we assume $\Delta F_c/F_c = 0.60$ or $\Delta m_c \approx 1.0$ to be the largest brightness variation\footnote{This value is based on the deepest eclipses found among 50099 EBs in the  All Sky Automated Survey catalogue \citep{ASUS-EB06}} that may give rise to a transit-like event, then a star has to fulfill the condition
\begin{equation}
m_{\mathit{diff}} \le -2.5\log\left((\Delta F/F)_s \,k^{-1}\right) - 0.55
\label{eq:maxconf}
\end{equation}
in order to be a potential contaminating source. This equation shows, for example, that a fully confused star with $k_c=1$ may not be more than 4.45 magnitudes fainter than the target in order to explain an observed variation of $(\Delta F/F)_s = 1/100$.

An optimised observing strategy should therefore first check if the neighborhood of a target contains any {\it potential false alarms}; that is, perform a search for contaminating stars that fulfill Eq.~\ref{eq:maxconf}. Only if such stars are found, is there any need to determine their brightness variations  in order to test the false alarm condition with Eqs.~\ref{eq:facond} to \ref{eq:fa_mag}.


\subsection{'On-off photometry' for transit follow-up}
{For the required identification of the contaminants' variability it is sufficient to compare their brightness during the central 'on' part of a predicted transit against the off-transit one, without a need to observe full transit events. Such 'on - off photometry' greatly reduces the observing time over that needed for the coverage of full transit events. Furthermore, 'on' and 'off' observations can -- and should -- be acquired at different nights, preverably with similar airmasses, allowing a more reliable identification of stellar variabilities in the case of transit events with long in- or egress durations, which else might be mistaken for extinction effects. We note that the continuous CoRoT lightcurve tells us if a target is stable from night to night or not, allowing to assess the reliability of on and off-observations taken on different nights.

For the \corot\ follow-up, we obtain short timeseries lasting from several minutes to about one hour for both the 'on' and the 'off' observations. Timeseries are preferable over single images for two reasons: for one, the brightness of \corot\ targets allows only exposure times up to $\approx 1$ min with the telescopes employed, and second, the scatter of the time-series points allows the estimation of measurement errors. 

The extraction of stellar brightnesses in 'on-off photometry' can in principle be done with any stellar photometry code. Here we employ mostly the PSF - deconvolution code {\it decphot} \citep{2007DECPHOT} -- best for cases were very close contaminants are apparent -- and the optimised aperture photometry package {\it vaphot} \citep{2001phot.work} for contaminants with PSF's that are separated in the ground-based images. For the latter one, the analyser program {\it vanaliz.pro} has been developed, which is described in the next section.

\subsection{{\it 'Vaphot/vanaliz'}, a timeseries photometry package with an analysis of on-off photometry}
{\it Vanaliz} is a universal program for the analysis of timeseries of a limited number of stars, geared towards the extraction of target lightcurves of eclipsing or transiting systems. Tables with stellar photometry, obtained from the associated IRAF routine are red as input, though adaptation to other formats would be straightforward. {\it Vanaliz} performs differential photometry against user-selectable sets of comparison stars, and contains routines for the interactive evaluation and weighting of comparison stars, for the output of statistical analyses and for manual or automatic suppression of data-points. While these routines have already been used in several projects for differential photometry (e.g. \citealt{deeg+00, martin+01, osorio+04, steindfadt+08, deeg+08, rabus+09}), a module that derives observed on-off brightness differences has been added for the \corot\ followup. Furthermore, it includes a calculation of the contamination factors $k_c$ from a simulated \corot\ PSF,  leading to estimations of the expected on-off brightness-variations of all contaminating stars. Comparing the observed with the expected on-off variations allows then the identification of false alarms soures; see Fig~\ref{fig:0165onoff} for an example. \footnote{A public version of this program without the simulation of the \corot\-PSF, using default values of  $k_c=1$, is available as the 'vaphot/vanaliz' package at\\ http://www.iac.es/galeria/hdeeg/hdeeghome.html}}

\section{The \corot\ photometric follow-up}

\subsection{The \corot photometric follow-up team}
Previous to \corot's launch we expected about 50 - 100 transit candidates per \corot\ observing run which will require further follow-up observations \citep{borde+03, garrido+06}. The expected need of observing and manpower resources led to the early planning of a dedicated photometric follow-up effort. Table~\ref{tab:instrum}  shows the instruments that have been used for the follow-up to date. They are mostly 1m-class instruments, which have proven sufficient for the large majority of follow-up observations. Observations with high angular resolution, eventually needed to identify contaminants very close to the target, have to date not been part of the regular follow-up program; they have been performed in only one case, towards the discovery of \exoseven\ b, and are described in \citet{leger+09}.

\begin{table}[!ht]
\begin{center}
\caption{Instruments used for \corot\ photometric follow-up}
\label{tab:instrum}
\scriptsize{
\begin{tabular}{l l}
\noalign{\smallskip}
\hline\hline
Instrument&Observatory\\
\hline
BEST 0.2-m$^a$&  Observatoire de Haute Provence, France\\
BESTII 0.25-m$^a$&  Cerro Armazones, Chile\\
TEST 0.3-m&Th\"uringer Landessternwarte, Germany\\
Vienna 0.8-m& Vienna Observatory, Austria\\
IAC 80-cm& Observatorio del Teide, Canary Islands, Spain\\
WISE 0.46-m, 1-m& WISE Observatory, Israel\\
OHP 1.2-m& Observatoire de Haute Provence, France\\
Euler 1.2-m telescope&La Silla, Chile\\
Tautenburg  2-m &Th\"uringer Landessternwarte, Germany\\
CFHT/Megacam 3.6-m &Mauna Kea, Hawaii\\
\hline
\hline
\end{tabular}
}
\end{center}
\scriptsize{$^a$ The small BEST and BEST II telescopes are not used in the follow-up as decribed here. They obtained previous timeseries of \corot's sample fields \citep{kabath+07, karoff+07, kabath+08, kabath+lrc2, kabath+lra2, rauer+09} which are routinely checked against the detections from \corot. They will also be used for future surveying of \corot-short run fields.}\\

\normalsize
\rm
\end{table}

\subsection{Observing strategy}

Planet candidates that enter the follow-up program are the result of the application of several different detection algorithms and the passing of a series of tests to classify the results \citep{moutou+05,moutou+07,almenara+07, barge+09}, which leads either to their rejection as false alarms or to their classification into one of several priority classes for planet candidates, according to their likeliness to be real planets. Observational follow-up is then only performed  on sufficiently highly classified candidates.

For the photometric follow-up of a candidate, we initially check for the existence of contaminating stars using the stellar photometry in the Exo-Dat catalog \citep{Deleuil+09}. A simulation of the \corot\ PSF allows then an estimation of the fraction of  light from the target and from the contaminating stars that falls into the target's aperture, giving the factors $k_c$. The evaluation based on Exo-Dat may be complemented by off-transit images from target fields taken with any of the telescopes participating in the follow-up effort. In cases where close blends are suspected, high-resolution imaging from adaptive optics or speckle imaging may also being obtained to aid in the identification of contaminants. Around the majority of candidates, these evaluations identify contaminating stars that are potential false alarms and which consequently require time-series follow-up photometry to derive their variability. Obtaining and analysing these time-series, part of which needs to be taken during candidates' transits, constitutes the major effort of the follow-up program. 

For the  assigning of observations and for the tracking of their results, we developed a web-based tool, the 'transit predictor'. It allows participating observers to obtain a list of the candidates that undergo transits at any night at their observatory, indicating their priority, transit start and end-times and corresponding target altitudes, moon-distances and solar altitudes, as well as general target information (position, brightness, transit depth). Since the number of transit candidates is fairly large, candidate transits that are in need of observation can be found for almost every night. This tool thus allows the optimum use of observing time when it is available only in nightly blocks within regular semestral scheduling; at most instruments this is still the only mode in which significant amounts of observing time can be obtained. Only for transit-events of  long-periodic cases with few observing opportunities, or for high priority objects (which may also need a specific instrument) do we intend to obtain observing time on short notice at specified moments. Since several groups are participating, a central coordination keeps track of the status of observations, reductions and their results, and accordingly prioritises the candidates, aiming to resolve the highest priorities rapidly while avoiding duplicate observations.

\begin{table*}[!ht]
\begin{center}
\caption{Potential false alarm sources around the E2\_0165 target star}
\label{tab:0165contam}
\scriptsize{
\begin{tabular}{lccccccc}
\noalign{\smallskip}
\hline\hline
Name & $\Delta$ RA & $\Delta$ DEC & m$_{\rm R}$  & k&relat. contrib. & observed&alarm  \\
 	& arcsec 		&	arcsec	&			&	&to target flux & $\Delta m_c$& $\Delta m_c$ \\
\hline
E2\_0165&0	  &0		&11.26	&1.00	&0.997&	0.021$\pm$0.004&0.00038\\
     c1&	8.3	&15.3	&15.33	  &0.10	        &0.0024& 0.023$\pm$0.033&0.17\\
     c2 &	16.3&-26.3	&13.50	&0.005	      &0.00069&0.011$\pm$0.018&0.76\\
c3	&2.4	&10.5	&20$^1$	&0.87	&0.00048	&$<$0.5$^2$&1.0\\

\hline

\end{tabular}
}\\

\scriptsize{$^1$g-band magnitude. $^2$Estimated  from movie made from CFHT time-series images.}\\
\end{center}
\normalsize
\rm
\end{table*}

\begin{figure}
   \centering
      \includegraphics[width=9cm]{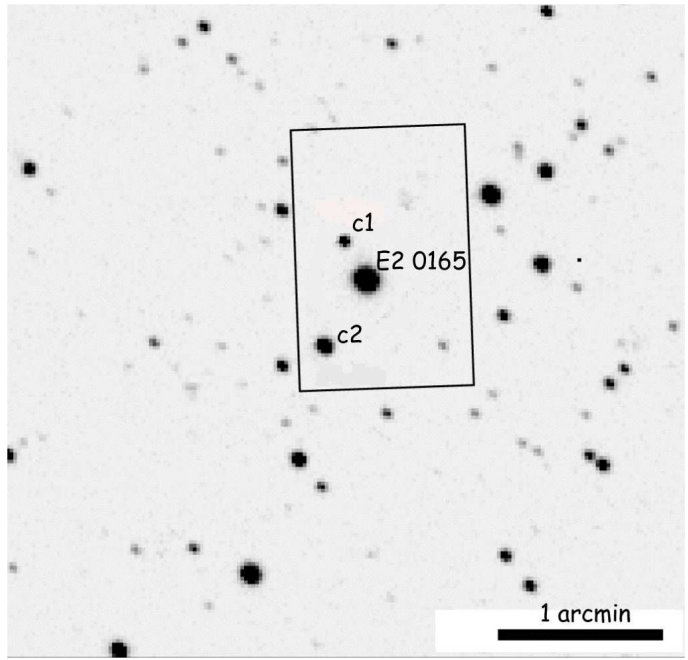}
  
   \caption{Finderchart of the area around target E2\_0165 made from an INT/WFC r-band image. N is to the top and E to the left. The rectangle shows the region of the 'imagette' of Fig.~\ref{fig:0165imagette} }
              \label{fig:0165fchart}
    \end{figure}

\begin{figure}
   \centering
      \includegraphics[width=9cm]{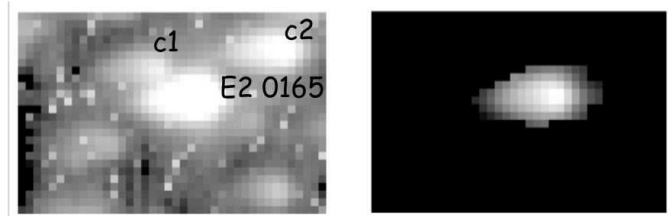}

   \caption{Imagette from Corot's CCD, centered on E2\_0165; orientation is N to the left, E to top. The right image shows the target through the aperture mask used for on-board photometry.}
              \label{fig:0165imagette}
    \end{figure}

\begin{figure}
   \centering
      \includegraphics[width=9cm]{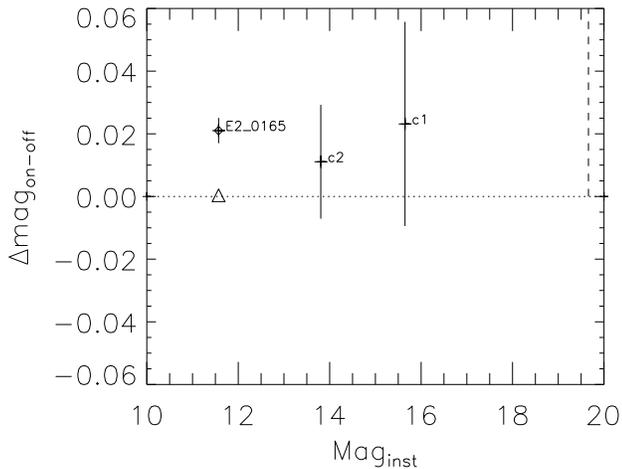}
  
   \caption{Difference between on and off-transit magnitudes for the target E2\_0165 and the two stars c1 and c2 that were potential false alarm sources near E2\_0165, based on the observations taken with the IAC80. In both cases the observed differences are much lower than required to be false alarm sources (see Table~\ref{tab:0165contam}) and are compatible with no variation. {The large observed offset for the E2\_0165 target of 0.02mag arises most likely from stellar activity and measurement errors; see text}. The triangle indicates its expected brightness variation of 0.4mmag -- too low to be resolved in this plot -- and the vertical dashed line indicates the approximate faint limit for false alarm sources, derived from Eq.~\ref{eq:maxconf}. Magnitudes are on an instrumental scale, 0.31mag larger than the m$_{\rm R}$ values of Table~\ref{tab:0165contam}}
              \label{fig:0165onoff}
    \end{figure}

\begin{figure}
   \centering
      \includegraphics[width=8cm]{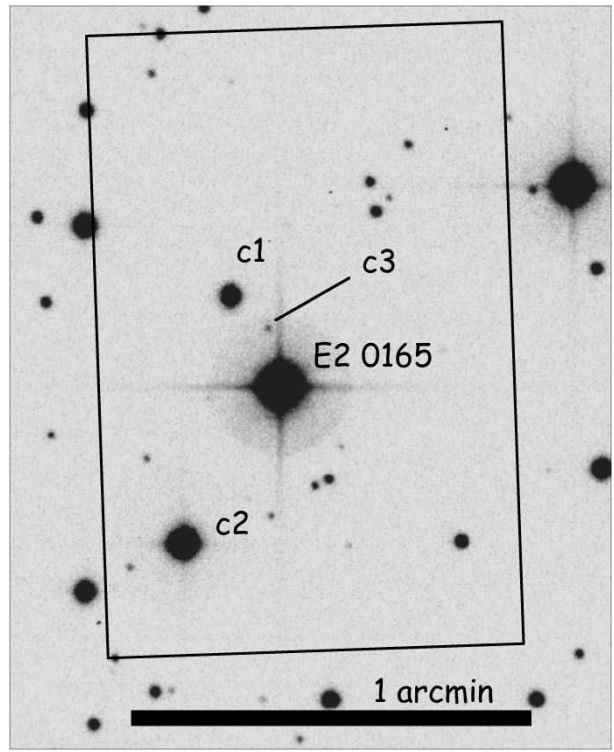}
  
   \caption{Image of E2\_0165 field taken with CFHT/Megacam, showing also faint contaminant c3.  N is to the top and E to the left and the rectangle corresponds again to  Fig.~\ref{fig:0165imagette}}
              \label{fig:0165cfht}
    \end{figure}

\subsection{An example of follow-up photometry: the exclusion of contaminants near the \exoseven\ host star}
As an example of the methodology of the \corot\ photometric follow-up, we show here in more detail the sequence that led to the exclusion of contaminants near the target star around which the terrestrial planet  \exoseven\ b has been detected. A summary of this section has been included in \citet{leger+09}, in which also all the other follow-up observations towards this discovery are being described. 

In a first step, once a new candidate is included in the follow-up program, a finderchart (Fig~\ref{fig:0165fchart}) is generated, based on Sloan-r filter images that had been obtained on the 2.5m INT/WFC instrument as part of the mission preparation. The Exo-Dat database which contains photometry extracted from the same images, also includes an estimation of the contamination level of a given target based on catalogued nearby stars and employing a generic \corot\ PSF. For \exoseven, which corresponds to Corot-ID 102708694 or target number LRa01\_E2\_0165, Exo-Dat indicates a low contamination level of 0.00053; meaning that about 0.05\% of the light detected from that target may come from contaminating stars.

  \begin{figure}
   \centering
    \includegraphics[width=9cm]{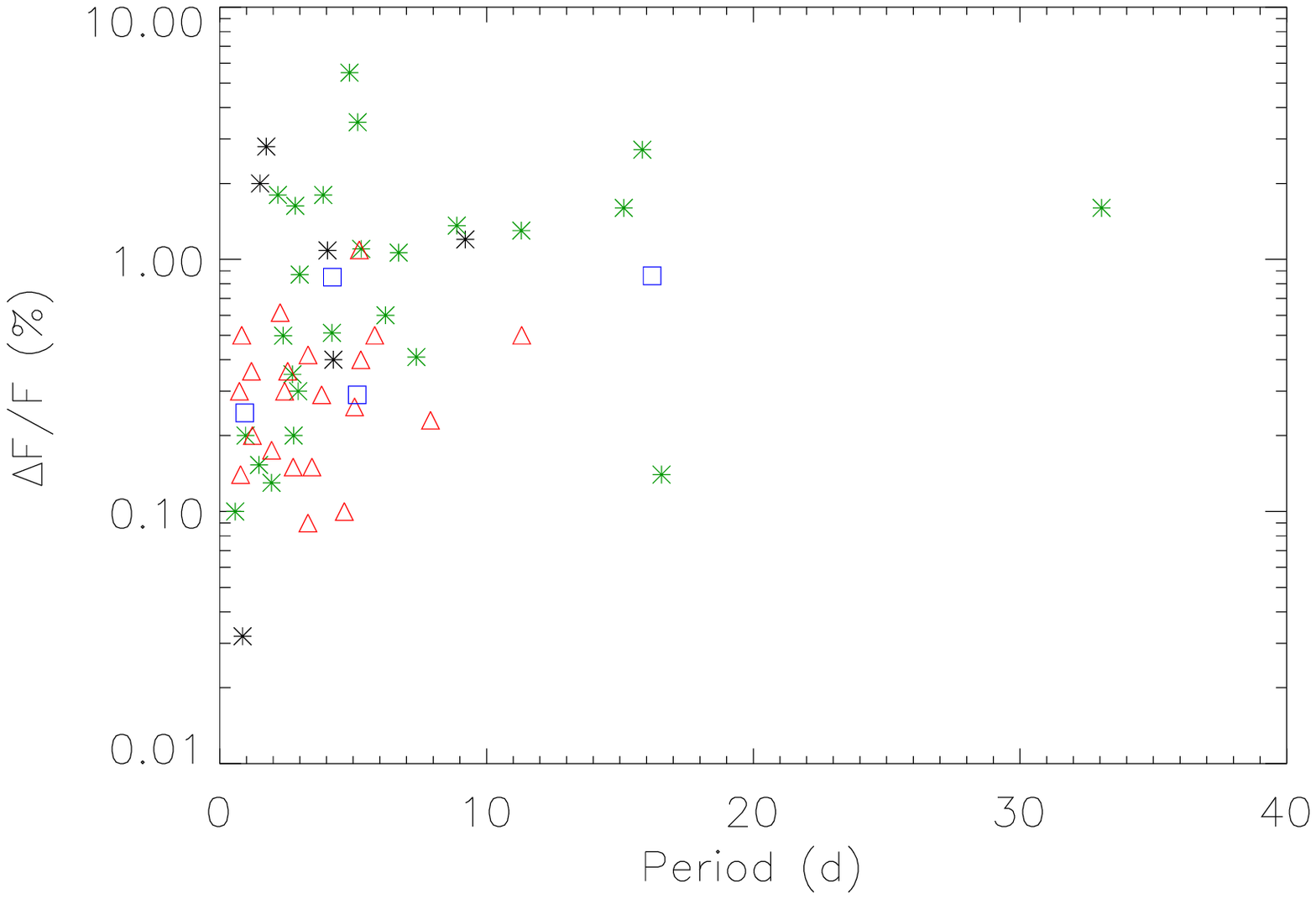}
    \includegraphics[width=9cm]{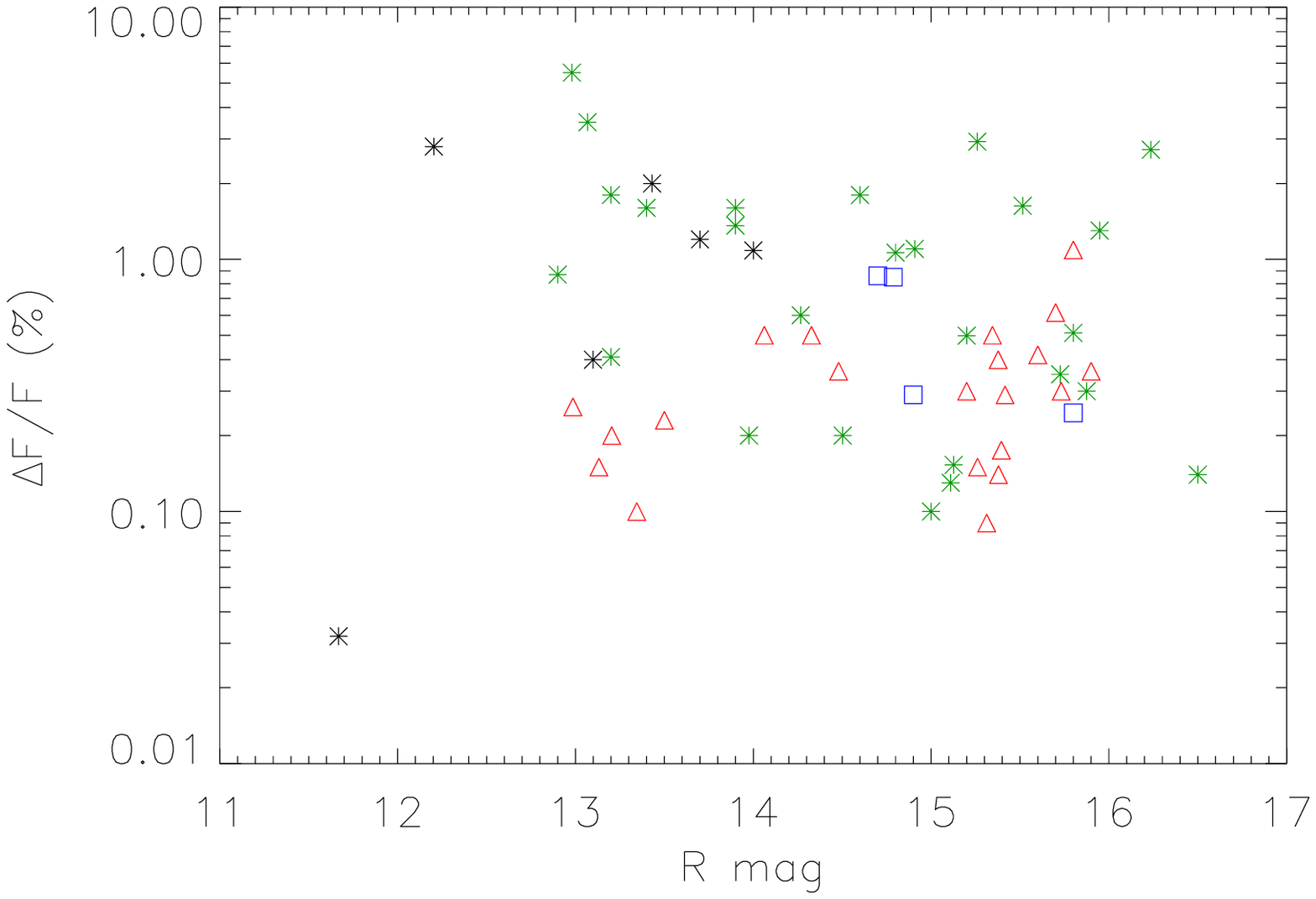}
  
   \caption{\corot\ planet candidates that were surveyed by the photometric follow-up, showing the candidate transit depth versus period (upper panel) and target magnitude (lower panel). Systems found by the photometric-follow up to be on-target are indicated by star-symbols, where the six published planets CoRoT-1b to 5b and 7b are differentiated by the black color (Corot-1b to 4b have previously been published under the name CoRoT-Exo-1b to 4b). CEBs that were detected by the follow-up are indicated by red triangles, whereas blue squares indicate cases were no clear result could be obtained.}
              \label{fig:df_vs_p.ps}
    \end{figure}

\begin{figure}
   \centering
      \includegraphics[width=9cm]{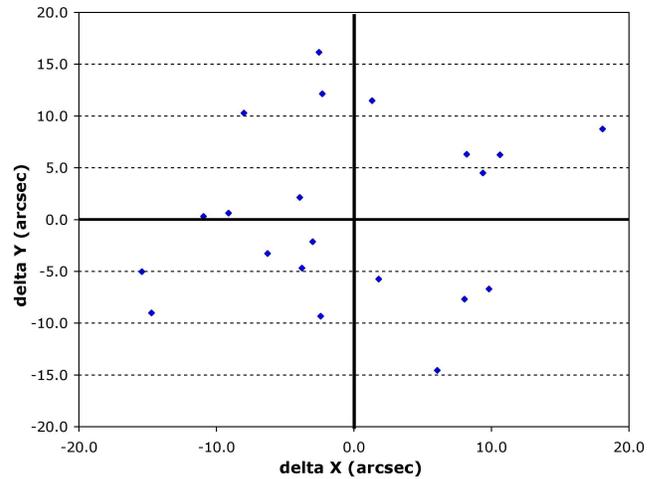}
  
   \caption{Distances of detected CEB from their target stars, set at positon (0,0). Distances are given in arcsec along the \corot\ CCD's X-Y axes. For the pointings towards the galactic center, these axes correspond approximately (ignoring small roll-angles of the satellite) to: N to the right, E downwards; for the anticenter pointings they are:  N to the left, E upwards.} 
              \label{fig:canddist}
    \end{figure}

\begin{figure}
   \centering
      \includegraphics[width=9cm]{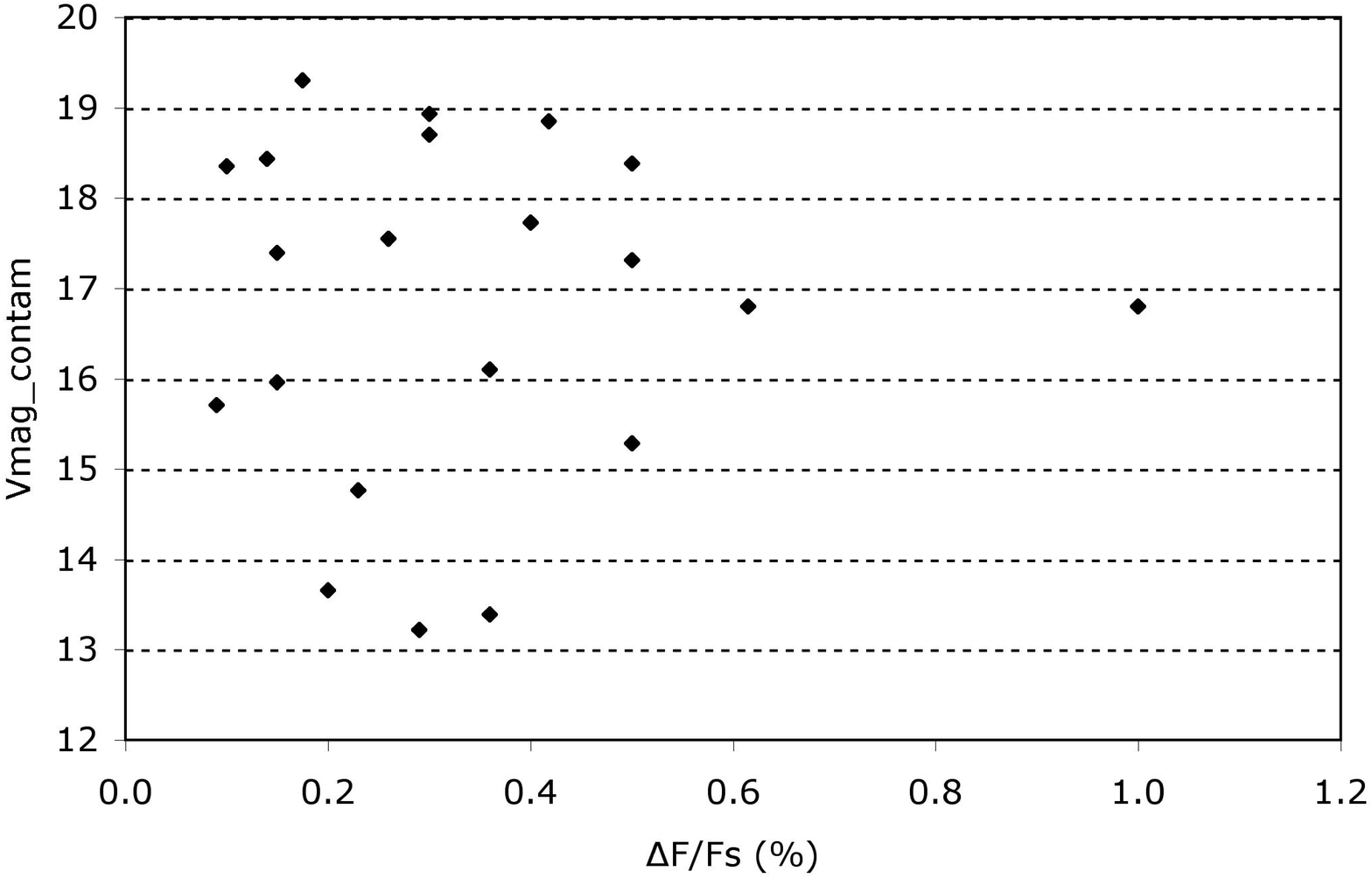}
      \includegraphics[width=9cm]{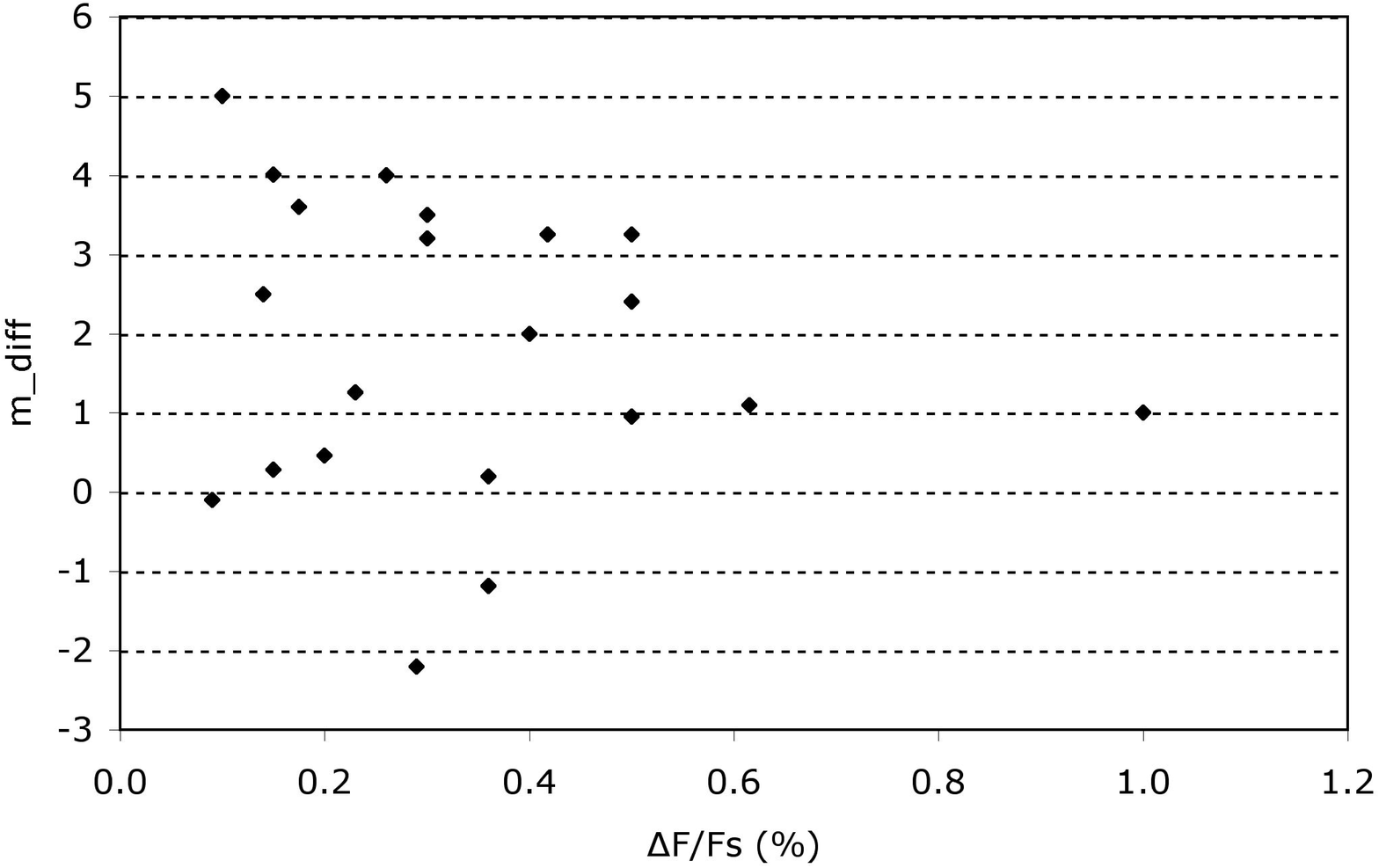}

   \caption{Top panel: Brightness of identified contaminating EBs versus the amplitude of eclipses in the discovery signal from CoRoT, $(\Delta F/F)_s$. Bottom panel: Magnitude difference (contaminant -- target) versus $(\Delta F/F)_s$. }
              \label{fig:cebs}
    \end{figure}

Follow-up observations were initially performed with the IAC80 telescope; a 40 minute long timeseries with a cadence of 40 seconds was taken in R filter during a transit on 27 Feb. 2008, and similar  off-transit observations were obtained on 5 Mar and 14 Apr 2008.  A model of  \corot's PSF of target E2\_0165 was then obtained from an extraction of its stellar image from a \corot\ - 'imagette'  (Fig.~\ref{fig:0165imagette}), a small subsection of an image from the \corot\ CCD around the target.  Considering the shape of \corot's target mask (right panel of Fig.~\ref{fig:0165imagette}) and using stellar brightnesses from an off-transit image taken at IAC80, this allowed a more precise calculation of the brightness variations that would be expected from any CEB. Based on this, two 'critical' nearby stars, \emph{c1} of R=15.3 and \emph{c2} of R=13.5 mag, were identified as potential false alarm sources, requiring eclipses with alarm amplitudes of 0.17 and 0.76 mag respectively (Table~\ref{tab:0165contam}). We must emphasize that all other nearby stars, not listed in Table~\ref{tab:0165contam}, were also tested to determine if they could be potential false alarm sources, but they could be excluded because they did not fulfill Eq.~\ref{eq:maxconftheo}, requiring impossibly large brightness variations.

 \emph{vaphot} was used to extract differential photometry and \emph{vanaliz} was used to calculate the differences between on- and off-transit fluxes, $\Delta$ mag$_{\rm{on-off}}$, for the target and the stars \emph{c1} and  \emph{c2}. This analysis (Fig.~\ref{fig:0165onoff}) showed that their on-off differences remained significantly below the required signal for a false-alarm (Table~\ref{tab:0165contam}).  It should be noted that an on-off difference of about 0.02 mag was found for the target; this is however much larger than the amplitude from \corot\ data and cannot be taken as a verification of the target as the alarm-source; {it is most likely due to a combination of  stellar activity, which causes brightness variations on the percent level over spans of several days \citep{leger+09} and of measurement errors}. Star \emph{c2} was furthermore identified as a separate \corot\ target (E2\_2180) with a lightcurve excluding any relevant brightness variations.

Further time-series images were taken using \textsc{CFHT/MEGACAM} \citep{2003SPIE.4841...72B} during a  transit-ingress on 7 March 2008. One exposure of 10 sec in the g-band was taken about each minute during 100 
minutes.  Only one CCD among the 40 chips of MEGACAM was actually used for that, which was 
sufficient, given a local star density that offered a wealth of nearby reference stars 
within the proper magnitude range.The image quality was medium for Mauna Kea (seeing $\approx$ 0.9 to 1.3 arcsec) but good enough for our purposes. These images, much deeper than those from IAC80, showed several additional faint nearby stars (Fig~\ref{fig:0165cfht}); however, only the star  \emph{c3} of $\approx$ 20 mag and 10\arcsec  NNE of the target was marginally admittable as a potential false alarm source, requiring unusually deep eclipses with amplitudes $\ga$1 mag. This star was visually inspected by generating a movie from the time-series images, indicating no apparent brightness variation. The contaminants c1 and c2 as well as all further nearby stars - though not critical -  were also demonstrated by the CFHT time-series to exhibit no significant brightness variations.

In conclusion, all stars around E2\_0165 could be excluded as false alarm sources. The remaining possiblity were very close CEBs that may have become diluted in the PSF of the target on the CFHT images. A further search (described in more detail in \citealt{leger+09}) for the presence of yet-unseen very close companions brighter than $\approx$ 20 mag (or $m_{\mathit{diff}} < 8.5$, from Eq.~\ref{eq:maxconftheo}, where $k\approx 1$ due to the closeness to the target) was then initiated using several high-resolution imaging techniques. That gave a negative result; indicating that alarm sources can be excluded at all distances larger than about 1 arcsec.

\subsection{Results from the photometric follow-up}

\begin{table*}[!ht]
\begin{center}
\caption{Photometric on-target identifications of \corot\ detections}
\label{tab:ontarget}
\scriptsize{
\begin{tabular}{lcccc|ccccc|c|l}
\noalign{\smallskip}
\hline\hline
&&candidate&&&observation&observed&&&alarm&observatory&comment\\
COROT\_ID&RUN-ID&$(\Delta F/F)_s$ (\%)&period&m$_t$&method&$(\Delta F/F)_t (\%)$&m$_c$&m$_{\mathit{diff}}$&$\Delta m_c$&&\\
\hline
102912369&IRa01\_E1\_0330&1.2&9.2&13.7&TV&1.3& &&&WISE, Euler, CFHT&Corot-4b\\
102829121&IRa01\_E1\_0399&1.6&33.06&13.9&TV&1.6& &&&WISE&\\
102779966&IRa01\_E1\_4108&0.41&7.37&15.4&CVA&&19.3&3.9&0.25&CFHT&\\
102825481&IRa01\_E2\_0203&3&5.17&13.3&TV&3.5& &&&IAC, CFHT&\\
102890318&IRa01\_E2\_1126&2&1.51&13.6&TV&2& &&&WISE, CFHT, BEST$^1$&Corot-1b\\
102826302&IRa01\_E2\_1712&0.2&2.77&14.0&CVA&&18.2&4.2&0.28&WISE, CFHT&\\
102954464&IRa01\_E2\_3856&0.14&16.56&16.5&TV&1.6& &&&Euler&\\
102764809&LRa01\_E1\_1031&1.09&4.04&14.0&CVA&&17.3&3.3&0.5&Vienna, IAC&Corot-5b\\
102708694&LRa01\_E2\_0165&0.03&0.85&11.3&CVA&&15.3&4.1&0.17&CFHT, IAC&Corot-7b$^2$\\
102615551&LRa01\_E2\_1123&1.8&3.88&14.6&TV&1.7& &&&IAC, WISE&\\
102716818&LRa01\_E2\_3156&0.15&1.47&15.8&NC&& &&&IAC&\\
102671819&LRa01\_E2\_3459&1.63&2.83&15.5&NC&& &&&IAC&\\
102755764&LRa01\_E2\_3739&2.93&61.48&15.6&TV&2.5& &&&Euler&=IRa01\_E1\_4014\\
102725122&LRa01\_E2\_5277&0.35&2.72&16.1&NC&& &&&IAC&\\
102582529&LRa01\_E2\_5756&2.72&15.84&16.2&TV&3& &&&IAC&\\
101065348&LRc01\_E1\_0499&5.5&4.86&13.3&TV&6& &&&IAC&\\
101368192&LRc01\_E1\_0523&0.4&4.26&13.3&TV&0.5& &&&IAC&Corot-3b\\
101206989&LRc01\_E1\_1929&0.1&0.58&15.0&CVA&&20.0&5&0.11&IAC&Low significance$^3$\\
101055792&LRc01\_E1\_2140&0.13&1.94&15.1&CVA&&17.2&2.1&0.02&IAC&\\
101644400&LRc01\_E1\_3268&1.1&5.3&15.3&NC&& &&&IAC, Vienna&\\
101206560&LRc01\_E2\_0192&2.8&1.74&12.6&TV&2.9& &&&WISE, IAC, TLS, BEST$^4$&Corot-2b\\
101086161&LRc01\_E2\_1145&0.6&6.21&14.8&TV&0.7& &&&OHP&\\
100768215&LRc01\_E2\_4390&0.3&2.94&16.5&CVA&&20.4&3.9&0.12&CFHT, IAC&\\
100834293&LRc01\_E2\_5414&1.3&11.3&16.4&CVA&&18.4&2&0.14&IAC&\\
105833549&LRc02\_E1\_0202&0.87&2.99&12.9&TV&1.2& &&&Euler&\\
106017681&LRc02\_E1\_0632&1.36&8.89&13.9&TV&1.4& &&&Euler, WISE&\\
105859159&LRc02\_E1\_0801&0.2&0.97&14.5&TV,CVA&0.5&17.7&3.2&0.04&TTB&Low significance$^5$\\
105416004&LRc02\_E1\_1427&0.5&2.38&15.2&TV&0.5& &&&Euler&TV 1.5hr early$^6$\\
105874733&LRc02\_E2\_1207&1.06&6.71&14.8&TV&1.2& &&&TLS&\\
105209106&LRc02\_E2\_4747&0.51&4.21&15.8&TV$^7$, CVA&0.8&16.7&0.9&0.11&Euler, IAC\\
221656539&SRa02\_E2\_0486&1.6&15.14&13.4&CVA&&13.6&0.2&0.45&WISE\\
221613770&SRa02\_E2\_0749&1.8&2.18&13.2&TV&1.8&&&&WISE\\
\hline
\end{tabular}
}\\
\end{center}
\scriptsize{$^1$Archival detection in pre-launch obs from 9 Dec 2006; $^2$See also Table 2; $^3$On-target classification on 2-sigma level;  $^4$Several archival detections in pre-launch obs from summer 2005; $^5$Low-confidence result (TV detection marginal and only 4\% variations in contaminant needed); $^6$TV was detected 1.5hr earlier than predicted from \corot\, requiring an ephemeris error on a 2-sigma level; $^7$TV not detected by IAC and unreliably by Euler, corrobated with CVA}\\

\normalsize
\rm
\end{table*}

\begin{table*}[!ht]
\begin{center}
\caption{Contaminating eclipsing binary (CEB) identifications around \corot\ targets}
\label{tab:ceb}
\scriptsize{
\begin{tabular}{lcccc|ccccc|c}
\noalign{\smallskip}
\hline\hline

&&candidate&&&&&contaminator&&& observatory\\
COROT-ID&RUN-ID&$(\Delta F/F)_s$ (\%)&period&m$_t$&m$_c$&$m_{\mathit{diff}}$&$(\Delta F/F)_c$(\%)&distance (arcsec)&k&\\
\hline
102787048&IRa01\_E1\_0288&0.23&7.90&13.50&14.76&1.26&3.0&9.1&0.24&OHP\\
102809071&IRa01\_E2\_1136 &0.2&1.22&13.20&13.66&0.46&8.0&12.4&0.038&Euler, Wise\\
102798247&IRa01\_E2\_1857 &0.5&0.82&14.06&17.31&3.25&11.5&3.7&0.87&Euler\\
102805893&IRa01\_E2\_2604 &0.29&3.82&15.42&13.22&-2.2&2.5&16.3&0.015&Euler\\
102835817&IRa01\_E2\_3425&0.36&1.19&15.9&16.1&0.20&26&13.0&0.017&Euler\\
102802430&IRa01\_E2\_4300 &0.5&5.81&14.33&15.28&0.95&51&10.4&0.024&CHFT\\
102714746&LRa01\_E1\_0544&0.15&2.75&13.39&17.40&4.0&20&9.6&0.30&Euler, IAC\\
102618931&LRa01\_E1\_2890 &0.3&2.43&15.73&18.93&3.2&40&12.3&0.14&IAC\\
102692038&LRa01\_E1\_4353 &1.0&5.23&15.80&16.80&1.0&4.5&11.1&0.56&Euler, IAC\\
102590008&LRa01\_E2\_4129&0.175&1.94&15.71&19.31&3.6&7.0&4.5&0.69&WISE, Euler\\
101095286&LRc01\_E1\_2376 &0.26&5.05&13.55&17.55&4.0&10.5&6.0&0.99&WISE, IAC\\
101436549&LRc01\_E1\_3129 &0.3&0.73&15.20&18.70&3.5&47.5&11.9&0.16&IAC\\
101175376&LRc01\_E1\_4959 &0.4&5.29&15.73&17.73&2.0&42.5&20.1&0.059&IAC\\
101106246&LRc01\_E2\_3257 &0.5&11.32&15.98&18.38&2.4&8.0&10.9&0.57&Euler, IAC\\
110604224&LRc01\_E2\_3681 &0.15&3.45&15.68&15.96&0.28&15.0&11.5&0.013&Euler, IAC\\
100609705&LRc01\_E2\_3895 &0.09&3.30&15.81&15.71&-0.1&16.8&15.8&0.005&Euler, IAC\\
100589010&LRc01\_E2\_4006 &0.14&0.78&15.93&18.43&2.5&35&16.2&0.040&WISE\\
211666030&SRc01\_E1\_3835 &0.62&2.26&15.70&16.80&1.1&5.0&10.3&0.34&IAC\\
110830906&LRa02\_E2\_2689 & 0.36&2.55&14.84&13.39&-1.18&28&17.3&0.004&BEST II\\
105681575&LRc02\_E1\_0483 &0.1&4.67&13.35&18.35&5.0&17.0&7.1&0.59&WISE\\
105574879&LRc02\_E2\_3142 &0.42&3.31&15.60&18.85&3.25&15.3&6.0&0.56&Euler, IAC\\

\hline

\end{tabular}
}\\

\scriptsize{.}\\
\end{center}
\normalsize
\rm
\end{table*}

A detailed reporting of the general characteristics of the non-planetary \corot\ candidates will be  the subject of a paper forthcoming at the end of the mission. First general results on the types of false alarm sources are given in \citet{almenara+09}, whereas detailed descriptions of the candidates of the first long runs are being published by 
 \citet{ira01+09} for the IRa01; by \citet{lrc01+09} for the LRc01 and by \citet{lra01+09} for the LRa01; with publications on later runs planned once a run's follow-up operations are reasonably complete. Here we limit ourselves to give an overview of the results from the photometric follow-up of the candidates that have been analysed through March 2009.  Fig.~\ref{fig:df_vs_p.ps} shows the sample of the surveyed cases against their period and magnitude. We note that the distribution of the surveyed candidates in these diagrams does not differ significantly from the distributions of the entire sample of planet candidates that have been selected for follow-up \citep{barge+09}. {An overall fraction of about 40\% of the candidates with pending photometric follow-up has been observed, with a higher rate of 80\% for the first \corot\ pointings (IRa01, LRc01); the remaining candidates are still pending observation}. Of the 57 candidates that have been surveyed by the \corot photometric follow-up team, 32 were found to be on-target, and 21 -- or about 35\% -- were background variables. 4 additional cases could not be resolved, either due to serious doubts about the candidate's ephemeris (from which it became doubtful that on-transit observations were done at the correct moment) or due to ambiguities in the photometry. Only one system (\exoseven) with a brightness variation significantly below 0.1\% has been surveyed; this is due to the sparseness of such planet candidates from \corot's lightcurves. 
{ Table~\ref{tab:ontarget} gives an overview about the candidates whose brightness variation was found to be on-target. Its leftmost five columns give basic information about the candidates from \corot's measurements. Next, the method is indicated by which the 'on-target' status was determined by photometric follow-up: TV (target variation) -- a brightness-variation of correct amplitude was found on the target-star itself; CVA (contaminant variations absent) --  the absence of any sufficiently strong variation in any contaminating star has been verified; NC (no contaminant) --  no contaminating star was found that could be a potential false alarm source.  For cases with observed target variations (TV), the next column indicates the observed target amplitude $(\Delta F/F)_t $. On the other hand, for cases without contaminant variations (CVA), $m_c$, the contaminant brightness, $m_\mathit{diff}$, its brightness difference to the target, and $\Delta m_c$,  the contaminant's brightness variation that is required to indicate an alarm is given. If more then one contaminant was present around a  target, we indicate the values for the contaminant with the smallest required alarm-amplitude $\Delta m_c$. The observed precision of contaminant brightness-variations can generally be taken to be 3\% or better; hence most cases identified in Table~\ref{tab:ontarget} are reliable identifications; those cases were only identifications of low reliability could be obtained are indicated by comments in the table. For the cases that were identified by a direct observation of target brightness variations (TV), we note that the observed brightness variation $(\Delta F/F)_t $ is typically larger than the one found by \corot, $(\Delta F/F)_s $, something that can be expected due to \corot's dilution of the candidate signal in its PSF by the contaminants that are present. Since the photometric follow-up is only an intermediate step in the chain to verify planet candidates, all candidates listed in Table~\ref{tab:ontarget} are subject to further observations. This is typically done through radial velocity measurements, and we indicate only the identified planets. For the first \corot\ runs, the outcome of these observations is described in the aforementioned papers on individual runs; whereas for the later runs, the final outcome is in many cases still open.}

Table~\ref{tab:ceb} indicates the candidates that photometry discarded by identifying CEBs near the \corot\ targets. The contamination fractions $k$ were calculated from the observed contaminant eclipse amplitude and $m_{\mathit{diff}}$. We note that all detected background variables appeared on \corot-candidates with brightness variations of $(\Delta F/F)_s \la 1\%$ (Fig.~\ref{fig:df_vs_p.ps}). This is likely due to a more reliable classification of the brighter candidates that enter the follow-up program, for which cases with CEBs have been rejected with more reliability from the discovery lightcurves. Contaminating EBs can in general be found within a distance of $\approx$17 arcsec; their distance and direction is shown in Fig.~\ref{fig:canddist}. Due to the changing orientation of the \corot\ cameras, this distance is plotted in the coordinates of the CCD's X and y axes, in which the dispersion direction of \corot's PSF remains constant. As expected, the distribution of detected CEBs outlines the generic PSF shape, with the more distant cases arising from bright or deeply eclipsing CEBs. 

Fig.~\ref{fig:cebs} shows the brightness of the identiÞed CEB and the magnitude differences between contaminant and target, plotted against the discovery eclipse amplitude $(\Delta F/F)_s$ in both cases. While the bulk of the CEB are fainter than the target by 1-3.5 mag, several bright CEB whose signal leaks into the \corot\ photometric apertures have also been identified,  all of them being cases where the contaminant was relatively far away from the target with only a very small fraction of its light falling into the target aperture. To date, no CEB fainter than 19 mag, or 5 mag fainter than the target has been identified; limits that will certainly become extended when more low-amplitude candidates become available, with subsequent stronger instrumental demands on the follow-up photometry. With more cases of \corot\ candidates being resolved by photometric follow-up, it may also be expected that differences in stellar densities between the center and anticenter pointings reveal different fractions of candidates turning out to be CEBs, due to the different stellar densities in these fields, with more CEB's expected in the denser pointings towards the galactic center.
  
\section{Conclusion}

The ongoing follow-up program of \corot\ detections demonstrates that  ground-based photometric follow-up is an important step in the cadence that leads from the detection in the space mission's data to the announcement of a planet discovery. An efficient observational method is the observation of short sections during predicted on and off-transit phases. The ground-based follow-up of \corot\ has found that approximately a third of \corot\'s transit candidates are caused by contaminating background binary stars at target distances of 2 - 15". Ground-based follow-up is able to detect or reject contaminating eclipsing binaries even in very low amplitude cases, in which the transit-like signals detected by the space mission are impossible to reproduce from ground. This is due to the much higher angular resolution obtainable from the ground in comparison to \corot, even with modest ground-based instrumentation. This ability of ground-based photometry is especially important in cases of suspected low-mass planets, were a positive verification with radial-velocity measurements is impossible or very difficult. Radial velocity observations are neither able to detect contaminating eclipsing binaries when these are sufficiently distant from the target star to remain outside of a spectrograph's entry slit -- a domain where photometric follow-up excells. The two follow-up techniques therefore complement each other very well in the identification of false alarm sources. With the push by \corot\, \emph{Kepler} and further planned missions towards the detection of smaller planets, and implicitly, lower mass planets, the independent positive verification of planet detections will become increasingly difficult. Planet discoveries from transits will therefore increasingly become based on the rejection of all possible false-alarm sources. This is a game were ground-based photometry has shown its value, as is demonstrated by its r\^{o}le in the recent discovery of \exoseven b.

\begin{acknowledgements}
The Austrian team thanks ASA for funding the \corot\ project. The team at IAC acknowledges support by grants ESP2004-03855-C03-03 and ESP2007-65480-C02-02 of the Spanish Ministerio de Ciencia e Inovaci\'{o}n. The team at TLS acknowledges the support of DLR grants 50 OW 0204 and Deutsche Forschungsgemeinschaft Grant  Ei 409/14-1. Tsevi Mazeh acknowledges support from the Israeli Science Foundation, through grant nr. 655/07.

Some of the data published in this article were acquired with the IAC80 telescope operated by the Instituto de Astrof\'\i sica de Canarias in the Observatorio del Teide and special thanks is given to its staff for performing a large fraction of these observations. This research has made use of the Exo-Dat database, operated at LAM-OAMP, Marseilles, France, on behalf of the CoRoT/Exoplanet program, and whose input catalogue was made possible thanks to observations collected for years at the Isaac Newton Telescope (INT), operated on the island of La Palma by the Isaac Newton group in the Spanish Observatorio del Roque de Los Muchachos of the Instituto de Astrof\'\i sica de Canarias. This research has made use of the Exo-Dat database, operated at LAM-OAMP, Marseille, France, on behalf of the CoRoT/Exoplanet program. We thank the referee for his comments that improved the presentation of this paper.


\end{acknowledgements}


\bibliographystyle{aa} 
\bibliography{../../../../HJDmain}

\begin{thebibliography}{38}
\expandafter\ifx\csname natexlab\endcsname\relax\def\natexlab#1{#1}\fi

\bibitem[{{Aigrain} {et~al.}(2008){Aigrain}, {Collier Cameron}, {Ollivier},
  {Pont}, {Jorda}, {Almenara}, {Alonso}, {Barge}, {Bord{\'e}}, {Bouchy},
  {Deeg}, {de La Reza}, {Deleuil}, {Dvorak}, {Erikson}, {Fridlund}, {Gondoin},
  {Gillon}, {Guillot}, {Hatzes}, {Lammer}, {Lanza}, {L{\'e}ger}, {Llebaria},
  {Magain}, {Mazeh}, {Moutou}, {Paetzold}, {Pinte}, {Queloz}, {Rauer}, {Rouan},
  {Schneider}, {Wuchter}, \& {Zucker}}]{aigrain+08}
{Aigrain}, S., {Collier Cameron}, A., {Ollivier}, M., {et~al.} 2008, \aap, 488,
  L43

\bibitem[{{Almenara} {et~al.}(2009){Almenara}, {Deeg}, {Aigrain}, {Alonso},
  {Barbieri}, {Bord{\'e}}, {Bouchy}, {Deeg}, {La Reza}, {Deleuil}, {Dvorak},
  {Erikson}, {Fridlund}, {Gillon}, {Gondoin}, {Guillot}, {Hatzes}, {Hebrard},
  {Jorda}, {Kabath}, {Lammer}, {Llebaria}, {Loeillet}, {Magain}, {Mazeh},
  {Moutou}, {Ollivier}, {P{\"a}tzold}, {Queloz}, {Rouan}, {Shporer}, \&
  {Wuchterl}}]{almenara+09}
{Almenara}, J., {Deeg}, H., {Aigrain}, S., {et~al.} 2009, submitted

\bibitem[{{Almenara} {et~al.}(2007){Almenara}, {Deeg}, {R{\'e}gulo}, \&
  {Alonso}}]{almenara+07}
{Almenara}, J.~M., {Deeg}, H.~J., {R{\'e}gulo}, C., \& {Alonso}, R. 2007, in
  Astronomical Society of the Pacific Conference Series, Vol. 366, Transiting
  Extrapolar Planets Workshop, ed. C.~{Afonso}, D.~{Weldrake}, \& T.~{Henning},
  183

\bibitem[{{Alonso} {et~al.}(2008){Alonso}, {Auvergne}, {Baglin}, {Ollivier},
  {Moutou}, {Rouan}, {Deeg}, {Aigrain}, {Almenara}, {Barbieri}, {Barge},
  {Benz}, {Bord{\'e}}, {Bouchy}, {de La Reza}, {Deleuil}, {Dvorak}, {Erikson},
  {Fridlund}, {Gillon}, {Gondoin}, {Guillot}, {Hatzes}, {H{\'e}brard},
  {Kabath}, {Jorda}, {Lammer}, {L{\'e}ger}, {Llebaria}, {Loeillet}, {Magain},
  {Mayor}, {Mazeh}, {P{\"a}tzold}, {Pepe}, {Pont}, {Queloz}, {Rauer},
  {Shporer}, {Schneider}, {Stecklum}, {Udry}, \& {Wuchterl}}]{alonso+08a}
{Alonso}, R., {Auvergne}, M., {Baglin}, A., {et~al.} 2008, \aap, 482, L21

\bibitem[{{Alonso} {et~al.}(2004){Alonso}, {Deeg}, {Brown}, \&
  {Belmonte}}]{adb+04}
{Alonso}, R., {Deeg}, H.~J., {Brown}, T.~M., \& {Belmonte}, J.~A. 2004,
  Astronomische Nachrichten, 325, 594

\bibitem[{{Baglin} {et~al.}(2007){Baglin}, {Auvergne}, {Barge}, {Michel},
  {Catala}, {Deleuil}, \& {Weiss}}]{baglin+07}
{Baglin}, A., {Auvergne}, M., {Barge}, P., {et~al.} 2007, in American Institute
  of Physics Conference Series, Vol. 895, American Institute of Physics
  Conference Series, ed. C.~{Dumitrache}, N.~A. {Popescu}, M.~D. {Suran}, \&
  V.~{Mioc}, 201--209

\bibitem[{{Barge} {et~al.}(2009){Barge}, {Almenara}, {Baglin}, {Auvergne},
  {Rauer}, {L{\'e}ger}, {Schneider}, {Pont}, {Aigrain}, {Almenara}, {Alonso},
  {Barbieri}, {Bord{\'e}}, {Bouchy}, {Deeg}, {La Reza}, {Deleuil}, {Dvorak},
  {Erikson}, {Fridlund}, {Gillon}, {Gondoin}, {Guillot}, {Hatzes}, {Hebrard},
  {Jorda}, {Kabath}, {Lammer}, {Llebaria}, {Loeillet}, {Magain}, {Mazeh},
  {Moutou}, {Ollivier}, {P{\"a}tzold}, {Queloz}, {Rouan}, {Shporer}, \&
  {Wuchterl}}]{barge+09}
{Barge}, P., {Almenara}, J.-M., {Baglin}, A., {et~al.} 2009, in prep.

\bibitem[{{Barge} {et~al.}(2008){Barge}, {Baglin}, {Auvergne}, {Rauer},
  {L{\'e}ger}, {Schneider}, {Pont}, {Aigrain}, {Almenara}, {Alonso},
  {Barbieri}, {Bord{\'e}}, {Bouchy}, {Deeg}, {La Reza}, {Deleuil}, {Dvorak},
  {Erikson}, {Fridlund}, {Gillon}, {Gondoin}, {Guillot}, {Hatzes}, {Hebrard},
  {Jorda}, {Kabath}, {Lammer}, {Llebaria}, {Loeillet}, {Magain}, {Mazeh},
  {Moutou}, {Ollivier}, {P{\"a}tzold}, {Queloz}, {Rouan}, {Shporer}, \&
  {Wuchterl}}]{barge+08}
{Barge}, P., {Baglin}, A., {Auvergne}, M., {et~al.} 2008, \aap, 482, L17

\bibitem[{{Bord{\'e}} {et~al.}(2003){Bord{\'e}}, {Rouan}, \&
  {L{\'e}ger}}]{borde+03}
{Bord{\'e}}, P., {Rouan}, D., \& {L{\'e}ger}, A. 2003, \aap, 405, 1137

\bibitem[{{Boulade} {et~al.}(2003){Boulade}, {Charlot}, {Abbon}, {Aune},
  {Borgeaud}, {Carton}, {Carty}, {Da Costa}, {Deschamps}, {Desforge},
  {Eppell{\'e}}, {Gallais}, {Gosset}, {Granelli}, {Gros}, {de Kat}, {Loiseau},
  {Ritou}, {Rouss{\'e}}, {Starzynski}, {Vignal}, \&
  {Vigroux}}]{2003SPIE.4841...72B}
{Boulade}, O., {Charlot}, X., {Abbon}, P., {et~al.} 2003, in Presented at the
  Society of Photo-Optical Instrumentation Engineers (SPIE) Conference, Vol.
  4841, Society of Photo-Optical Instrumentation Engineers (SPIE) Conference
  Series, ed. M.~{Iye} \& A.~F.~M. {Moorwood}, 72--81

\bibitem[{{Brown}(2003)}]{brown03}
{Brown}, T.~M. 2003, \apjl, 593, L125

\bibitem[{Cabrera {et~al.}(2009)Cabrera, Fridlund, Ollivier, Gandolfi,
  Csizmadia, Alonso, Aigrain, Alapini, \& et. al.}]{lrc01+09}
Cabrera, J.~., Fridlund, M., Ollivier, M., {et~al.} 2009, accepted

\bibitem[{Carone {et~al.}(2009)Carone, Hatzes, Cabrera, Gandolfi, Csizmadia,
  Alonso, Aigrain, Alapini, \& et~al.}]{lra01+09}
Carone, L., Hatzes, A., Cabrera, J., {et~al.} 2009, in prep.

\bibitem[{{Deeg} \& {Doyle}(2001)}]{2001phot.work}
{Deeg}, H.~J. \& {Doyle}, L.~R. 2001, in Third Workshop on Photometry, p. 85,
  ed. W.~J. {Borucki} \& L.~E. {Lasher}, 85

\bibitem[{{Deeg} {et~al.}(2000){Deeg}, {Doyle}, {Kozhevnikov}, {Blue},
  {Mart{\'{\i}}n}, \& {Schneider}}]{deeg+00}
{Deeg}, H.~J., {Doyle}, L.~R., {Kozhevnikov}, V.~P., {et~al.} 2000, \aap, 358,
  L5

\bibitem[{{Deeg} {et~al.}(2008){Deeg}, {Oca{\~n}a}, {Kozhevnikov},
  {Charbonneau}, {O'Donovan}, \& {Doyle}}]{deeg+08}
{Deeg}, H.~J., {Oca{\~n}a}, B., {Kozhevnikov}, V.~P., {et~al.} 2008, \aap, 480,
  563

\bibitem[{{Deleuil} {et~al.}(2008){Deleuil}, {Deeg}, {Alonso}, {Bouchy},
  {Rouan}, {Auvergne}, {Baglin}, {Aigrain}, {Almenara}, {Barbieri}, {Barge},
  {Bruntt}, {Bord{\'e}}, {Collier Cameron}, {Csizmadia}, {de La Reza},
  {Dvorak}, {Erikson}, {Fridlund}, {Gandolfi}, {Gillon}, {Guenther}, {Guillot},
  {Hatzes}, {H{\'e}brard}, {Jorda}, {Lammer}, {L{\'e}ger}, {Llebaria},
  {Loeillet}, {Mayor}, {Mazeh}, {Moutou}, {Ollivier}, {P{\"a}tzold}, {Pont},
  {Queloz}, {Rauer}, {Schneider}, {Shporer}, {Wuchterl}, \&
  {Zucker}}]{deleuil+08}
{Deleuil}, M., {Deeg}, H.~J., {Alonso}, R., {et~al.} 2008, \aap, 491, 889

\bibitem[{{Deleuil} {et~al.}(2009){Deleuil}, {Meunier}, {Moutou}, {Surace},
  {Deeg}, {Barbieri}, {Debosscher}, {Almenara}, \& {Granet}}]{Deleuil+09}
{Deleuil}, M., {Meunier}, J., {Moutou}, C., {et~al.} 2009, \aj, accepted

\bibitem[{{Deleuil} {et~al.}(2006){Deleuil}, {Moutou}, {Deeg}, {Meunier},
  {Guterman}, {Almenara}, {Alonso}, {Bouchy}, {Erikson}, \& {et.
  al}}]{Deleuil+06}
{Deleuil}, M., {Moutou}, C., {Deeg}, H., {et~al.} 2006, in ESA Special
  Publication, Vol. 1306, ESA Special Publication, 341

\bibitem[{{Eisl{\"o}ffel} {et~al.}(2004){Eisl{\"o}ffel}, {K{\"u}rster},
  {Hatzes}, \& {Guenther}}]{eisloeffel04}
{Eisl{\"o}ffel}, J., {K{\"u}rster}, M., {Hatzes}, A.~P., \& {Guenther}, E.
  2004, in ESA Special Publication, Vol. 538, Stellar Structure and Habitable
  Planet Finding, ed. F.~{Favata}, S.~{Aigrain}, \& A.~{Wilson}, 81

\bibitem[{{Fridlund} {et~al.}(2006){Fridlund}, {Baglin}, {Lochard}, \&
  {Conroy}}]{corotbook06}
{Fridlund}, M., {Baglin}, A., {Lochard}, J., \& {Conroy}, L., eds. 2006, "The
  CoRoT Mission" (ESA Publications Division, ESA Spec.Publ. 1306)

\bibitem[{{Garrido} \& {Deeg}(2006)}]{garrido+06}
{Garrido}, R. \& {Deeg}, H.~J. 2006, Lecture Notes and Essays in Astrophysics,
  vol.~2, p.~27-48., 2, 27

\bibitem[{{Gillon} {et~al.}(2007){Gillon}, {Magain}, {Chantry}, {Letawe},
  {Sohy}, {Courbin}, {Pont}, \& {Moutou}}]{2007DECPHOT}
{Gillon}, M., {Magain}, P., {Chantry}, V., {et~al.} 2007, in Astronomical
  Society of the Pacific Conference Series, Vol. 366, Transiting Extrapolar
  Planets Workshop, ed. C.~{Afonso}, D.~{Weldrake}, \& T.~{Henning}, 113

\bibitem[{{Kabath} {et~al.}(2007){Kabath}, {Eigm{\"u}ller}, {Erikson},
  {Hedelt}, {Rauer}, {Titz}, {Wiese}, \& {Karoff}}]{kabath+07}
{Kabath}, P., {Eigm{\"u}ller}, P., {Erikson}, A., {et~al.} 2007, \aj, 134, 1560

\bibitem[{{Kabath} {et~al.}(2008){Kabath}, {Eigm{\"u}ller}, {Erikson},
  {Hedelt}, {von Paris}, {Rauer}, {Renner}, {Titz}, \& {Karoff}}]{kabath+08}
{Kabath}, P., {Eigm{\"u}ller}, P., {Erikson}, A., {et~al.} 2008, \aj, 136, 654

\bibitem[{Kabath {et~al.}(2009)Kabath, Erikson, Rauer, Csizmadia, Chini, \&
  Lemke}]{kabath+lra2}
Kabath, P., Erikson, A., Rauer, H., {et~al.} 2009, \aap, submitted

\bibitem[{{Kabath} {et~al.}(2009){Kabath}, {Fruth}, {Rauer}, {Erikson},
  {Murphy}, {Chini}, {Lemke}, {Csizmadia}, {Eigm{\"u}ller}, {Pasternacki}, \&
  {Titz}}]{kabath+lrc2}
{Kabath}, P., {Fruth}, T., {Rauer}, H., {et~al.} 2009, \aj, 137, 3911

\bibitem[{{Karoff} {et~al.}(2007){Karoff}, {Rauer}, {Erikson}, {Voss},
  {Kabath}, {Wiese}, {Deleuil}, {Moutou}, {Meunier}, \& {Deeg}}]{karoff+07}
{Karoff}, C., {Rauer}, H., {Erikson}, A., {et~al.} 2007, \aj, 134, 766

\bibitem[{{L\'eger} {et~al.}(2009){L\'eger}, Rouan, Schneider, Alonso, Samuel,
  {Aigrain}, {Almenara}, {Alonso}, {Barbieri}, {Bord{\'e}}, {Bouchy}, {Deeg},
  {La Reza}, {Deleuil}, {Dvorak}, {Erikson}, {Fridlund}, {Gillon}, {Gondoin},
  {Guillot}, {Hatzes}, {Hebrard}, {Jorda}, {Kabath}, {Lammer}, {Llebaria},
  {Loeillet}, {Magain}, {Mazeh}, {Moutou}, {Ollivier}, {P{\"a}tzold}, {Queloz},
  {Rouan}, {Shporer}, \& {Wuchterl}}]{leger+09}
{L\'eger}, A., Rouan, D., Schneider, J., {et~al.} 2009, submitted

\bibitem[{{Mart{\'{\i}}n} {et~al.}(2001){Mart{\'{\i}}n}, {Zapatero Osorio}, \&
  {Lehto}}]{martin+01}
{Mart{\'{\i}}n}, E.~L., {Zapatero Osorio}, M.~R., \& {Lehto}, H.~J. 2001, \apj,
  557, 822

\bibitem[{{Moutou} {et~al.}(2007){Moutou}, {Aigrain}, {Almenara}, {Alonso},
  {Auvergne}, {Barge}, {Blouin}, {Borde}, {Cabrera}, {Carone}, {Cautain},
  {Deeg}, {Erikson}, {Fressin}, {Guis}, {Leger}, {Guterman}, {Irwin}, {Kabath},
  {Lanza}, {Maceroni}, {Mazeh}, {Ollivier}, {Pont}, {Paetzold}, {Queloz},
  {Rauer}, {Rouan}, {Schneider}, {Tamuz}, {Voss}, \& {Zucker}}]{moutou+07}
{Moutou}, C., {Aigrain}, S., {Almenara}, J., {et~al.} 2007, in Astronomical
  Society of the Pacific Conference Series, Vol. 366, Transiting Extrapolar
  Planets Workshop, ed. C.~{Afonso}, D.~{Weldrake}, \& T.~{Henning}, 127

\bibitem[{{Moutou} {et~al.}(2005){Moutou}, {Pont}, {Barge}, {Aigrain},
  {Auvergne}, {Blouin}, {Cautain}, {Erikson}, {Guis}, {Guterman}, {Irwin},
  {Lanza}, {Queloz}, {Rauer}, {Voss}, \& {Zucker}}]{moutou+05}
{Moutou}, C., {Pont}, F., {Barge}, P., {et~al.} 2005, \aap, 437, 355

\bibitem[{Moutou {et~al.}(2009)Moutou, Pont, Bouchy, Deleuil, Almenara, Alonso,
  Barbieri, Bruntt, \& et. al.}]{ira01+09}
Moutou, C., Pont, F., Bouchy, F., {et~al.} 2009, accepted

\bibitem[{{Osorio} {et~al.}(2004){Osorio}, {Caballero}, {Mart{\'{\i}}n},
  {B{\'e}jar}, \& {Rebolo}}]{osorio+04}
{Osorio}, M.~R.~Z., {Caballero}, J., {Mart{\'{\i}}n}, E.~L., {B{\'e}jar},
  V.~J.~S., \& {Rebolo}, R. 2004, \apss, 292, 673

\bibitem[{{Paczy{\'n}ski} {et~al.}(2006){Paczy{\'n}ski}, {Szczygie{\l}},
  {Pilecki}, \& {Pojma{\'n}ski}}]{ASUS-EB06}
{Paczy{\'n}ski}, B., {Szczygie{\l}}, D.~M., {Pilecki}, B., \& {Pojma{\'n}ski},
  G. 2006, \mnras, 368, 1311

\bibitem[{{Rabus} {et~al.}(2009){Rabus}, {Alonso}, {Belmonte}, {Deeg},
  {Gilliland}, {Almenara}, {Brown}, {Charbonneau}, \& {Mandushev}}]{rabus+09}
{Rabus}, M., {Alonso}, R., {Belmonte}, J.~A., {et~al.} 2009, \aap, 494, 391

\bibitem[{Rauer {et~al.}(2009)Rauer, Erikson, Hedelt, Boer, \&
  Carone}]{rauer+09}
Rauer, H., Erikson, A.and~Kabath, P., Hedelt, P., Boer, M., \& Carone, L. 2009,
  \aj, submitted

\bibitem[{{Steinfadt} {et~al.}(2008){Steinfadt}, {Bildsten}, {Ofek}, \&
  {Kulkarni}}]{steindfadt+08}
{Steinfadt}, J.~D.~R., {Bildsten}, L., {Ofek}, E.~O., \& {Kulkarni}, S.~R.
  2008, \pasp, 120, 1103

\end{thebibliography}
 
\end{document}